\documentstyle[prl,aps,floats,epsf,color]{revtex}
\begin{document}
\baselineskip=12pt
\def\be{\begin{equation}}
\def\ee{\end{equation}}
\def\bea{\begin{eqnarray}}
\def\eea{\end{eqnarray}}
\def\E{{\rm e}}
\def\bearst{\begin{eqnarray*}}
\def\eearst{\end{eqnarray*}}
\def\peleven{\parbox{11cm}}
\def\peffec{\peight{\bearst\eearst}\hfill\peleven}
\def\pspace{\peight{\bearst\eearst}\hfill}
\def\ptwelve{\parbox{12cm}}
\def\peight{\parbox{8mm}}
\twocolumn[\hsize\textwidth\columnwidth\hsize\csname@twocolumnfalse\endcsname

\title
{Power-law Parameterized Quintessence Model}
\author{ Sohrab Rahvar$^{1,2}$,  M. Sadegh Movahed$^{1,2}$}
\address{$^{1}$Department of Physics, Sharif University of
Technology, P.O.Box 11365--9161, Tehran, Iran}
\address{$^{2}$ Institute for Studies in theoretical Physics and Mathematics, P.O.Box 19395-5531,Tehran,
Iran}

\vskip 1cm

 \maketitle


\begin{abstract}
We introduce a power-law parameterized quintessence model for the
dark energy which accelerate universe at the low redshifts while
behaves as an ordinary matter for the early universe.
We construct a unique scalar potential for this parameterized
quintessence model. As the observational test, the Supernova Type
Ia (SNIa) Gold sample data, size of baryonic acoustic peak from
Sloan Digital Sky Survey (SDSS), the position of the acoustic peak
from the CMB observations and structure formation from the
$2$dFGRS survey are used to constrain the parameters of the
quintessence model. The best fit parameters indicates that the
equation of state of this model at the present time is less than
one $(w_0<-1)$ which violates the energy condition in General
Relativity. Finally we compare the age of old objects with age of
universe in this model.
\end{abstract}
\hspace{.3in}
\newpage
]
\section{Introduction}
Observations of the apparent luminosity and redshift of type Ia
supernovas (SNIa) provide the main evidence for the positive
accelerating expansion of the Universe \cite{ris,permul}. A
combined analysis of SNIa and the Cosmic Microwave Background
radiation (CMB) observations indicates that the dark energy filled
about $2/3$ of the total energy of the Universe and the remained
part is dark matter with a few percent in the form of Baryonic
matter from the Big Bang nucleo synthesis
\cite{bennett,peri,spe03}.

The "cosmological constant" is a possible solution for the
acceleration of the universe \cite{wein}. This constant term in
Einstein field equation can be regarded as an fluid with the
equation of state of $w=-1$. However, there are two problems with
the cosmological constant, namely the {\it fine-tuning} and the
{\it cosmic coincidence}. In the framework of quantum field
theory, the vacuum expectation value is $123$ order of magnitude
larger that the observed value of $10^{-47}$ GeV$^{4}$. The
absence of a fundamental mechanism which sets the cosmological
constant zero or very small value is the cosmological constant
problem. The second problem as the cosmic coincidence, states that
why are the energy densities of dark energy and dark matter nearly
equal today?

One of the solutions to this problem is a model with varying
cosmological constant decays from the beginning of the universe to
a small value at the present time. A non-dissipative minimally
coupled scalar field, so-called Quintessence model can play the
role of time varying cosmological constant
\cite{wet88,amen,peb88}. The ratio of energy density of this field
to the matter density increases slowly by the expansion of the universe
and after a while the dark energy becomes the dominated term of
energy-momentum tensor. One of the features of this model is the
variation of equation of state during the expansion of the
universe. Various Quintessence models as k-essence \cite{arm00},
tachyonic matter \cite{pad03}, Phantom \cite{cal02,cal03} and
Chaplygin gas \cite{kam01} provide various equation of states for
the dark energy
{\cite{cal03,arb05,wan00,per99,pag03,dor01,dor02,dor04}.

There are also phenomenological models, parameterize the equation
of state of dark energy in terms of redshift
\cite{che01,lin03,sel04}. For a dark energy with the equation of
state of $p_X=w_X\rho_X$, using the continuity equation, the
density of dark energy changes with the scale factor as:
\begin{equation}
\rho_X = \rho_{X}^{(0)}a^{-3(1+\bar{w}_X(a))},
\end{equation}
where $\bar{w}_X(a)$ is the mean of the equation of state in the
logarithmic scale:
\begin{equation}
\bar{w}_X(a) = \frac{\int w_X(a) d\ln(a)}{\int{d\ln(a)}}
\label{mean_w}
\end{equation}
The main aim of these models is to cure the fine-tuning problem of
the dark energy density by means that the ratio of dark energy
density to the matter density $(\rho_m \sim a^{-3})$,
\begin{equation}
\frac{\rho_X}{\rho_m} = \frac{\rho_X^{(0)}}{\rho_m^{(0)}}a^{-3\bar{w}_X(a)}
\label{r_density}
\end{equation}
approach to unity at the early universe in contrast to that of
cosmological constant.
Here we propose a simple power-law model for the mean value of
equation of state as:
\begin{equation}
\bar{w}_X(a) = w_0 a^{\alpha},
\end{equation}
which can remove the fine-tuning problem of the dark energy at the
early universe. This model is expressed with the two parameter of
$w_0$ (equation of state at the present time) and the exponent of
$\alpha$. The equation of state of this model according to the
definition of ${\bar w_X}(a)$ obtain as:
\begin{equation}
w(a) = w_0a^{\alpha}(1+\ln{a^{\alpha}})
\end{equation}
One of the advantages of this model is that for the small scale
factors in the range of $a<e^{-1/\alpha}$, the sign of the
equation of state will change to the positive value and the dark
energy will behave as an ordinary matter. Figure (\ref{fig_dr})
shows the ratio of dark energy to the matter densities as a
function of scale factor for various values of $\alpha$. For
$\alpha<1.3$ we have two times of domination of dark energy during
the history of universe: once for the early universe and the other
time at the lower red-shifts. However, since the sign of the
equation of state of dark energy at the early time is positive, it
will not accelerate universe and we will have only one phase of
the acceleration for the later times. Figure (\ref{acc}) shows the
deceleration parameter $(q=-\ddot{a}a/{\dot{a}^2})$ of the universe in
terms of scale factor $a$ for various values of $\alpha$.
Increasing the $\alpha$-exponent causes the universe to enter the
acceleration phase of universe at the later times but speed-up it
to enter the de Sitter phase faster (see Figure \ref{acc}).
\begin{figure}
\epsfxsize=9.5truecm\epsfbox{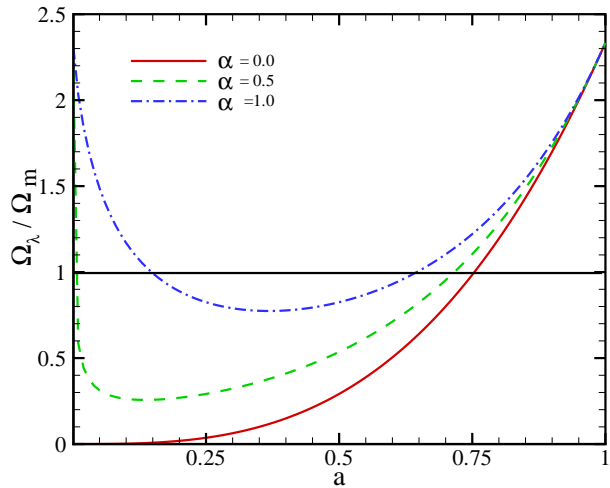} \narrowtext \caption{Ratio
of dark energy to the matter density as a function of scale
factor. For $0<\alpha<1.3$ we have two times of dark energy
dominance over the (cold dark) matter. It should be mentioned that
for the early universe the sign of equation of state will change to the positive value
and behaves as a cold dark matter. Here we choose $w_0 = -1$,
$\Omega_m=0.3$ and $\Omega_{tot}=1.0$.} \label{fig_dr}
\end{figure}

\begin{figure}
\epsfxsize=9.5truecm\epsfbox{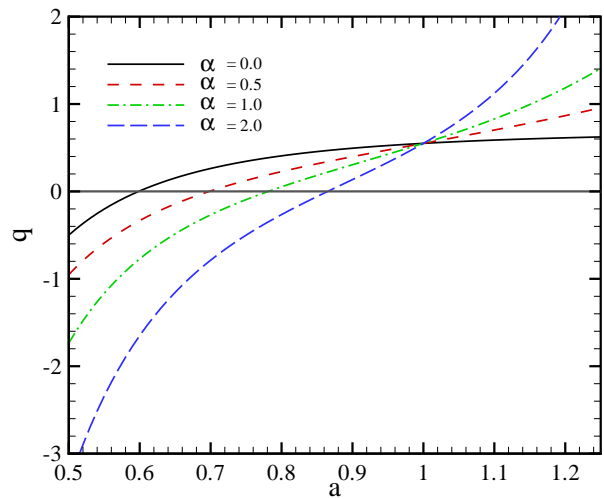} \narrowtext \caption{
Deceleration parameter $(q=-\ddot{a}a/{\dot{a}^2})$ in the
power-law model as a function of scale factor for various values
of $\alpha$-exponent. Increasing $\alpha$ causes that universe to
enter the acceleration phase at later times. } \label{acc}
\end{figure}

The organization of the paper is as follows: In Sec.\ref{scalar_f}
we reconstruct a scalar potential for generating the power-law
quintessence model. In Sec.\ref{lowredshift} we study the effect
of this model on the age of Universe, comoving distance, comoving
volume element and the variation of angular size by the redshift
\cite{alc79}. In Sec. \ref{cobs} we put constrain on the
parameters of model by the background evolution, such as Gold
sample of Supernova Type Ia data \cite{R04}, the position of the
observed acoustic angular scale on CMB and the baryonic
oscillation length scale. We study the linear structure formation
in this model and compare the growth index with the observations
from the $2-$degree Field Galaxy Redshift Survey ($2$dFGRS) data
in sec. \ref{cstructure}. We also compare the age of the universe
in this model with the age of old cosmological structures in this
section. Sec.\ref{conc} contains summary and conclusion of this
work.

\section{Corresponding potential of the scalar field }
\label{scalar_f}
Scalar field is one of the physical mechanism for
a time-varying dark energy that can fulfill the condition of
positive acceleration of the universe at the present time.
Essential condition for a given scalar field to play the role of
dark energy is that the equation of state at the lower redshifts
can provide the condition of $w<-1/3$. The energy density and
pressure of an homogeneous scalar field with the potential of
$V(\phi)$ and kinetic term of $\dot{\phi}^2/2$ are:
\begin{eqnarray}
\label{rho_eq}
\rho_X &=& \frac{1}{2}\dot{\phi}^2 + V(\phi) \\
P_X &=& \frac{1}{2}\dot{\phi}^2 - V(\phi) \label{pre_eq}
\end{eqnarray}
Using, the definition of equation of state of dark energy,
$w_X=P/\rho$, the equation of state in terms of kinetic and
potential energies of scalar field can be written as:
\begin{equation}
w_X = \frac{T+V}{T-V}, \label{e_state}
\end{equation}
The kinetic and potential energies of scalar field from the
equations ({\ref{rho_eq}) and (\ref{pre_eq}) in terms of
$\rho_X$ and the equation of state of dark energy are:
\begin{equation}
\label{kin}
T=\frac12\rho_X(1+w_X)
\end{equation}
\begin{equation}
V = \frac12\rho_X(1-w_X), \label{pot}
\end{equation}
For a positive $T$
and $V$ the equation of state is bounded to the interval
$-1<w<+1$. For $T>0$ and $V<0$ we have $|w|>1$ and for the case
of $T<0$ the equation of state can be $w<-1$ (i.e. kinetic term of
Lagrangian has negative sign). Here we reconstruct a scalar
potential which can generate the power-law quintessence model. The
kinetic term of the scalar field from the equation (\ref{kin})
in terms of redshift obtain as \cite{zong05}:
\begin{equation}
\frac{d \phi}{d z} = \pm
\frac{[\rho_X(z)(1+w_X)]^{1/2}}{H(z)(1+z)}, \label{dphi}
\end{equation}
where the minus or plus sign are chosen if $\dot{\phi}>0$ and
$\dot{\phi}<0$, respectively. Choosing the sign is arbitrary as it
can be changed by the field redefinition of $\phi \rightarrow
-\phi$. Here we choose the negative sign for convenient. The
Hubble parameter also is given by:
\begin{equation}
H^2(z)=H_0^2[\Omega^{(0)}_m(1+z)^3+\Omega^{(0)}_\lambda(1+z)^{3[1+\bar{w}(z)]}],
\label{eq4}
\end{equation}
where we can substitute the Hubble parameter at the present time
with $H_0^2 = \rho_c^{(0)}/{3 M_{pl}^2}$. Using the numerical
integration we obtain the dependence of the scalar field,
$\tilde{\phi}=\phi/M_{pl}$ in terms of the redshift (Figure
\ref{phi_redshift}). On the other hand using equation
(\ref{pot}) we can obtain the potential in terms of redshift as:
\begin{equation}
\tilde{V}(z) =\frac{1}{2}(1-w_X)F(z), \label{v_phi} \label{v_z}
\end{equation}
where $F(z) = \rho_X(z)/\rho_X(0)$ and 
$\tilde{V}(z)=V(z)/V(0)$. Finally we do numerical integration from
the equation (\ref{dphi}) and obtain the dependence of the
redshift to the scalar field $z = f(\phi)$. Substituting this
function in the equation (\ref{v_z}) results the explicit relation
between the potential and scalar field or in another word,
reconstruction of the scalar potential for the power-law
quintessence model is obtained (see Figure \ref{v_potential}).
\begin{figure}[t]
\epsfxsize=9.5truecm\epsfbox{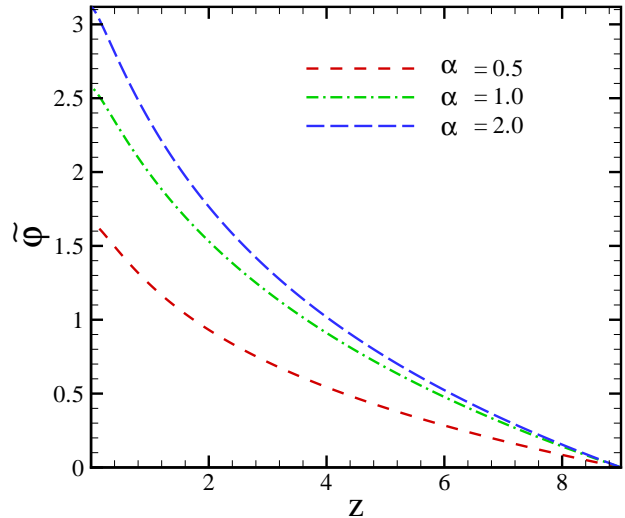} \narrowtext \caption{Dependence of scalar field in terms of redshift for the
power-law Quintessence model.}
 \label{phi_redshift}
\end{figure}
\begin{figure}[t]
\epsfxsize=9.5truecm\epsfbox{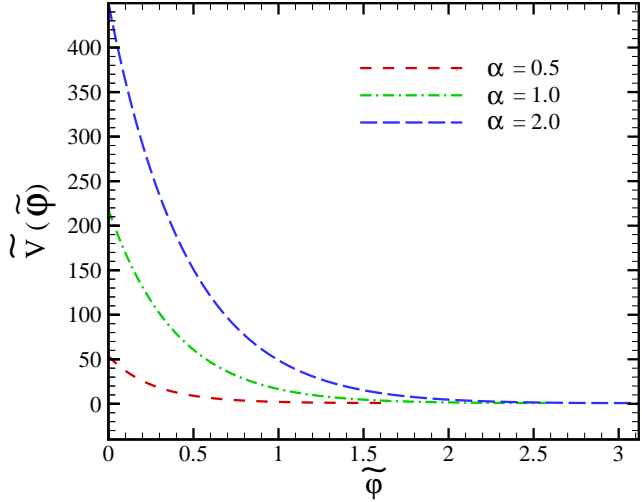} \narrowtext \caption{Reconstruction of the scalar field potential as a function of scalar field for the power-law Quintessence model.}
 \label{v_potential}
\end{figure}

\section{the effect of variable dark energy on the geometrical parameters of universe}
\label{lowredshift} The cosmological observations are mainly
affected by the background dynamics of universe. In this section
we study the effect of the power-law dark energy model on the
geometrical parameters of universe.

\subsection{comoving distance}
The radial comoving distance is one of the basis parameters of
cosmology. For an object with the redshift of $z$, using the null
geodesics in the FRW metric, the comoving distance obtain as:
\begin{equation}
r(z;\alpha,w_0)  =  \int_0^z {dz'\over H(z';\alpha,w_0)},
\label{comoving}
\end{equation}
where $H(z;\alpha,w_0)$ is the Hubble parameter and after the
matter-radiation equality  epoch, it can be expressed in terms of
Hubble parameter at the present time, $H_0$, matter and dark
energy content of the universe.

\begin{figure}[t]
\epsfxsize=9.5truecm\epsfbox{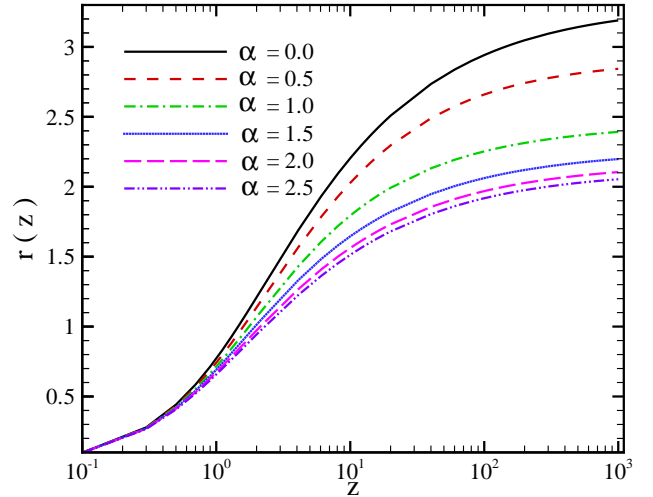} \narrowtext
\caption{Comoving distance, $r(z;\alpha,w_0)$ (in unit of $c/H_0$)
as a function of redshift for various values of $\alpha$. Here we
fix $w_0 = -1$.} \label{fig:rz}
 \end{figure}
By numerical integration of equation (\ref{comoving}), the
comoving distance in terms of redshift for different values of
$\alpha$ is shown in Figure~\ref{fig:rz}. Increasing the $\alpha$
exponent, increases the contribution of dark energy at the present
time and results a smaller comoving distance. One of the main
applications of the comoving distance calculation is on the
analyzing of luminosity distance of SNIa data.

\subsection{Angular Size}
Measurement of apparent angular size of an object located at the
cosmological distance is another important parameter that can be
affected by the amount and variation of dark energy during the
history of universe. An object with the physical size of $D$ is
related to the apparent angular size of $\theta$ by:
\begin{equation}
D=d_A \theta \label{as}
\end{equation}
where $d_A=r(z;\alpha,w_0)/(1+z)$ is the angular diameter
distance. The main applications of equation (\ref{as}) is on the
measurement of the apparent angular size of acoustic peak on CMB
and baryonic acoustic peak at the lower redshifts. By measuring
the angular size of an object in different redshifts (so-called
Alcock-Paczynski test) it is possible to probe the variability of
dark energy \cite{alc79}. The variation of apparent angular size
$\Delta\theta$ in terms $\Delta z$ is given by:
\begin{equation}
{\Delta z\over \Delta \theta} =
\frac{H(z;\alpha,w_0)r(z;\alpha,w_0)}{\theta} \label{alpa}
\end{equation}

\begin{figure}[t]
\epsfxsize=9.5truecm\epsfbox{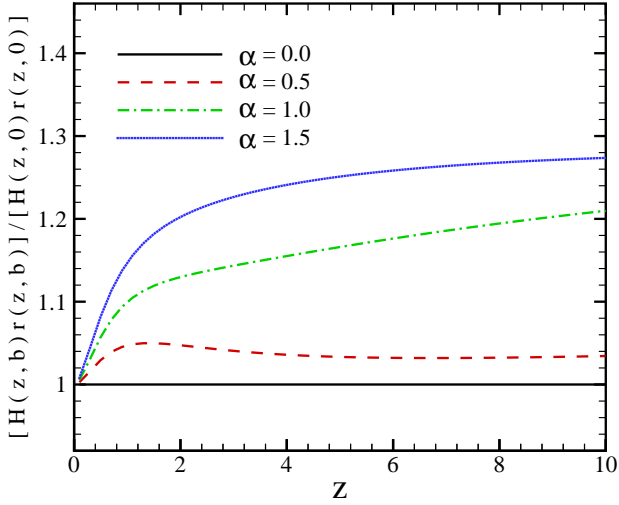} \narrowtext
\caption{Alcock-Paczynski test, compares $\Delta z/{\Delta
\theta}$ normalized to the case of $\Lambda$CDM model as a
function of redshift for four different $\alpha$. } \label{fig:hr}
 \end{figure}
Figure~\ref{fig:hr} shows $\Delta z/ \Delta \theta$ in terms of
redshift, normalized to the case with $\alpha=0$ (i.e.
$\Lambda$CDM model). The advantage of Alcock-Paczynski test is
that it is independent of standard candles and a standard ruler
such as the size of baryonic acoustic peak can be used to
constrain the dark energy model.
\subsection{Comoving Volume Element}
The comoving volume element is an other geometrical parameter
which is used in number-count tests such as lensed quasars,
galaxies, or clusters of galaxies. The comoving volume element in
terms of comoving distance and Hubble parameters is given by:
\begin{equation}
f(z;\alpha,w_0) \equiv {dV\over dz d\Omega} =
r^2(z;\alpha,w_0)/H(z;\alpha,w_0).
\end{equation}
According to Figure \ref{fig:v}, the comoving volume element
becomes maximum around $z\simeq 2$. For a larger $\alpha$
exponent, the position of the peak of comoving volume element
shifts to the lower redshifts.

\begin{figure}[t]
\epsfxsize=9.5truecm\epsfbox{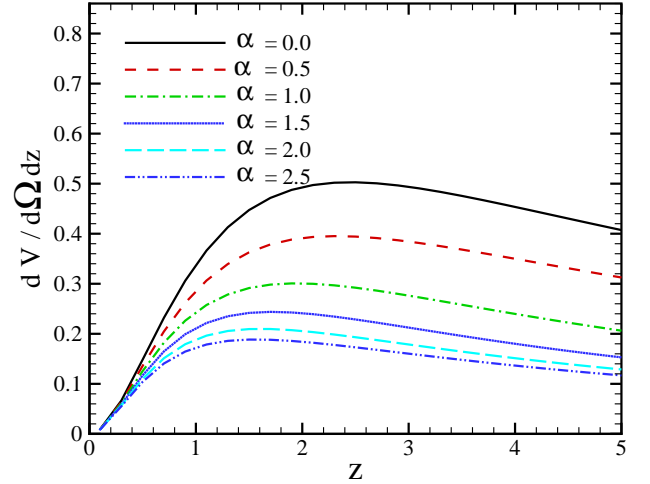} \narrowtext \caption{The
comoving volume element in terms of redshift for various $\alpha$
exponent. Increasing $\alpha$ shifts the position of maximum value
of volume element to the lower redshifts. } \label{fig:v}
 \end{figure}

\subsection{Age of Universe}
The "age crises" is one the main reasons for the existence of dark
energy. The problem is that the universe's age in the Cold Dark
Matter (CDM) universe is less than the age of old stars in it.
Studies on the old stars \cite{carretta00} suggests an age of
$13^{+4}_{-2}$ Gyr for the universe. Richer et. al.
\cite{richer02} and Hasen et. al. \cite{hansen02} also proposed an
age of $12.7\pm0.7$ Gyr, using the white dwarf cooling sequence
method (for full review of the cosmic age see \cite{spe03}). The
age of universe integrating from the big bang up to now obtain as:
\begin{equation}\label{age}
t_0(\alpha,w_0) = \int_0^{t_0}\,dt = \int_0^\infty {dz\over
(1+z)H(z;\alpha,w_0)},
\end{equation}
Figure~\ref{fig:1} shows the dependence of $H_0t_0$ (Hubble
parameters times the age of universe) on $\alpha$-exponent for a
typical values of cosmological parameters (e.g. $h=0.65$,
$\Omega_m=0.27$ and $w_0=-1.0$). Increasing $\alpha$ results a
shorter age for the universe.
\begin{figure}[t]
\epsfxsize=9.5truecm\epsfbox{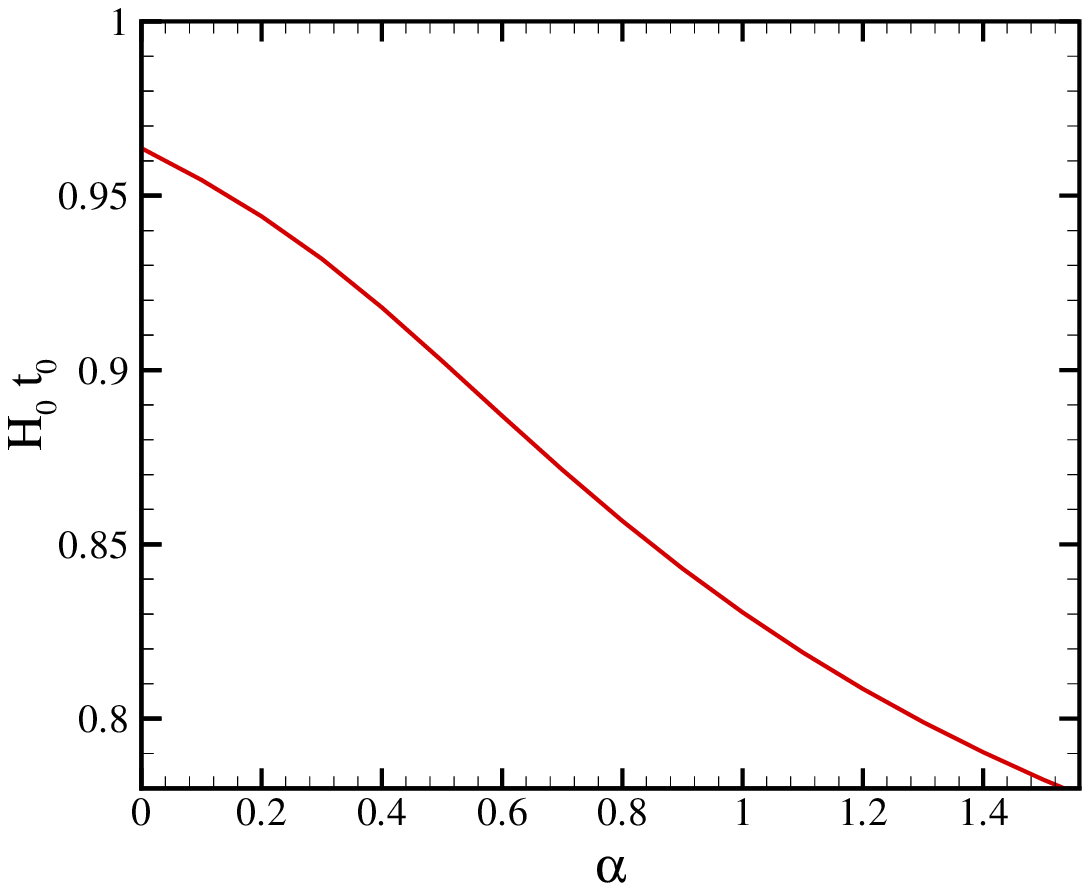} \narrowtext \caption{
$H_0t_0$ (age of universe times the Hubble constant at the present
time) as a function of $\alpha$ in a flat universe with the
parameters of $\Omega_{m}=0.3$, $h=0.65$ and $w_0=-1.0$.
Increasing $\alpha$-exponent makes a shorter age for the
universe.} \label{fig:1}
 \end{figure}


\section{Observational Constraint From the Background Evolution}
\label{cobs}
 In this section we compare the SNIa Gold sample data,
the location of baryonic acoustic peak from the SDSS and the
location of acoustic peak from the CMB observation to constrain
the parameters of model.

\subsection{Examining Model by Supernova Type Ia: Gold Sample}
\label{sn} The Supernova Type Ia experiments provided the main
evidence of the existence of dark energy. Since 1995 two teams of
the {\it High-Z Supernova Search} and the {\it Supernova Cosmology
Project} have been discovered several type Ia supernovas at the
high redshifts \cite{per99,Schmidt}. Recently Riess et al. ~(2004)
announced the discovery of $16$ type Ia supernova with the Hubble
Space Telescope. This new sample includes $6$ of the $7$ most
distant ($z> 1.25$) type Ia supernovas. They determined the
luminosity distance of these supernovas and with the previously
reported algorithms, obtained a uniform $157$ Gold sample of type
Ia supernovas \cite{R04,Tonry,bar04}.

 In this subsection we compare the
predictions of the dark energy model with the SNIa Gold sample. The
observations of supernova measure essentially the apparent magnitude
$m$ including reddening, K correction etc, which is related to the
(dimensionless) luminosity distance, $D_L$, of a an object at
redshift $z$, for a spatially flat universe by: \begin{equation}
m={\mathcal{M}}+5\log{D_{L}(z;\alpha,w_0)}, \label{m} \end{equation}
where
\begin{eqnarray}
\label{luminosity} D_L (z;\alpha,w_0) &=&
H_0(1+z)\int^{z}_{0}{d\zeta\over H(\zeta;\alpha,w_0)}\,.
\end{eqnarray}
Also
\begin{eqnarray}
\label{m1}\mathcal{M} &=& M+5\log{\left(\frac{c/H_0}{1\quad
Mpc}\right)}+25.
\end{eqnarray}
where $M$ is the absolute magnitude. The distance modulus, $\mu$, is
defined as:

\begin{equation}
\mu\equiv
m-M=5\log{D_{L}(z;\alpha,w_0)}+5\log{\left(\frac{c/H_0}{1\quad
Mpc}\right)}+25, \label{eq:mMr} \end{equation}

In order to compare the theoretical results with the observational
data, we must compute the distance modulus, as given by Eq.
(\ref{eq:mMr}). The first step in this sense is to compute the
quality of the fitting through the least squared fitting quantity
$\chi^2$ defined by:
\begin{eqnarray}\label{chi_sn}
\chi^2=\sum_{i}\frac{[\mu_{obs}(z_i)-\mu_{th}(z_i;\Omega_m,w_0,\alpha,h)]^2}{\sigma_i^2},
\end{eqnarray}
where $\sigma_i$ is the observational uncertainty in the distance
modulus. To constrain the parameters of model, we use the Likelihood
statistical analysis
 Marginalizing over the nuisance
parameter of $h$ in a flat universe $(\Omega_{total} =1)$, the
best fit values for the parameters of model obtain as
$w_0=-2.60_{-2.00}^{+1.80}$, $\Omega_m=0.45^{+0.09}_{-0.45}$ and
$\alpha=1.00_{-1.00}^{+1.00}$ with $\chi^2_{min}/N_{d.o.f} =1.13$
at $1 \sigma$ level of confidence. The corresponding value for the
Hubble parameter at the minimized $\chi^2$ is $h=0.66$ and since
we have already marginalized over this parameter we do not assign
an error bar for it. Figure \ref{modul1} shows the best fit of
model to the Gold sample of SNIa. We compare our result with that
of Riess et al. (2004), for $\alpha=0$ their result has been
recovered (see Figure \ref{fig61}).

For the age consistency test we substitute the parameters of model
from the SNIa fitting in equation (\ref{age}) and obtain the age
of universe about $13.19$ Gry, which is in good agreement with the
age of old stars.

\begin{figure}[t]
\epsfxsize=9.5truecm
\begin{center}
\epsfbox{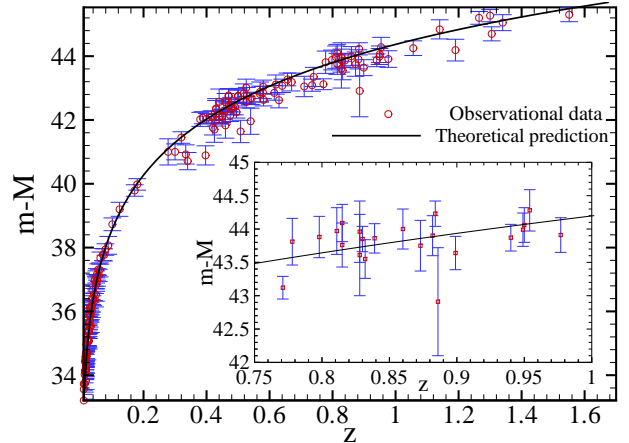} \narrowtext \caption{Fitting the distance
modulus of the SNIa Gold sample in terms of redshift with the
power-law dark energy model. Solid line shows the best fit with
the corresponding parameters of $h=0.66$,
$w_0=-2.60_{-2.00}^{+1.80}$, $\Omega_m^=0.45^{+0.09}_{-0.45}$ and
$\alpha=1.00_{+1.00}^{-1.00}$ in $1 \sigma$ level of confidence
with $\chi^2_{min}/N_{d.o.f} =1.13$} \label{modul1}
\end{center}
\end{figure}

\begin{figure}[t]
\begin{center}
\epsfxsize=9truecm\epsfbox{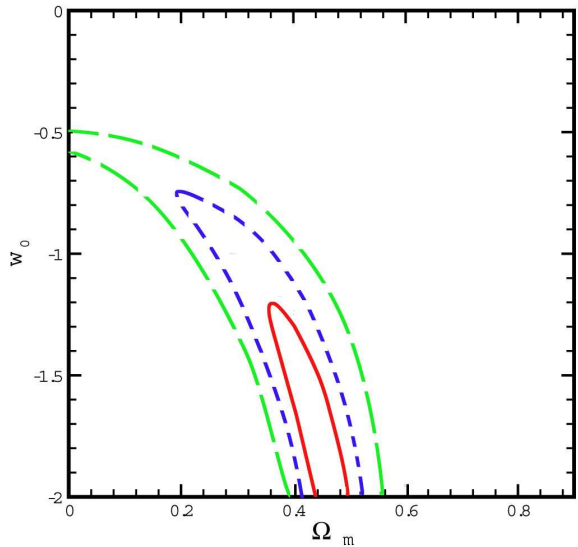} \narrowtext \caption { Joint
confidence intervals for $\Omega_m$ and $w_0$ for the case of
$\alpha=0$ with  $1\sigma$ (solid-line), $2\sigma$ (dashed-line)
and $3\sigma$ (long dashed-line) confidence level. This result is
in good agreement with that of Riess et al. (2004).} \label{fig61}
\end{center}
\end{figure}

\subsection{Combined analysis: SNIa$+$CMB$+$SDSS} \label{cmb}
In this section we combine SNIa Gold sample, CMB data from the
WMAP with recently observed baryonic peak from the SDSS to
constrain the parameters of power-law dark energy model
\cite{b03}.

The apparent acoustic peak is the most relevant parameter in the
spectrum of CMB which can be used to determine the geometry and
the matter content of universe. The acoustic peak corresponds to
the Jeans length of photon-baryon structures at the last
scattering surface some $\sim 379$ Kyr after the Big Bang
\cite{spe03}. The apparent angular size of acoustic peak in a flat
universe can be obtained by dividing the comoving size of sound
horizon at the decoupling epoch $r_s(z_{dec})$ to the comoving
distance of observer to the last scattering surface $r(z_{dec})$:
\begin{equation}
\theta_A \equiv {{r_s(z_{dec})}\over r(z_{dec}) }.
\label{eq:theta_s}
\end{equation}
The size of sound horizon at numerator of equation
(\ref{eq:theta_s}) corresponds to the a distance that a
perturbation of pressure can travel from the beginning of universe
up to the last scattering surface and obtain by:
\begin{equation}
r_s(z_{dec};\alpha,w_0) =
\int_{z_{dec}}^{\infty}\frac{v_s(z)}{H(z;\alpha,w_0)}dz,
\label{sh}
\end{equation}
where $v_s(z)^{-2}=3 + 9/4\times\rho_b(z)/\rho_r(z)$ is the sound
velocity in the unit of speed of light from the big bang up to the
last scattering surface \cite{dor01,Hu95}.

Changing the parameters of the dark energy can change the size of
apparent acoustic peak and subsequently the position of $l_A\equiv
\pi/\theta_A$ in the power spectrum of temperature fluctuations on
CMB. Here we plot the dependence of $l_A$ on $\alpha$ and $w_0$
for a typical values of cosmological parameters (see Figure
\ref{lab}).
\begin{figure}[t]
\begin{center}
\epsfxsize=9.5truecm\epsfbox{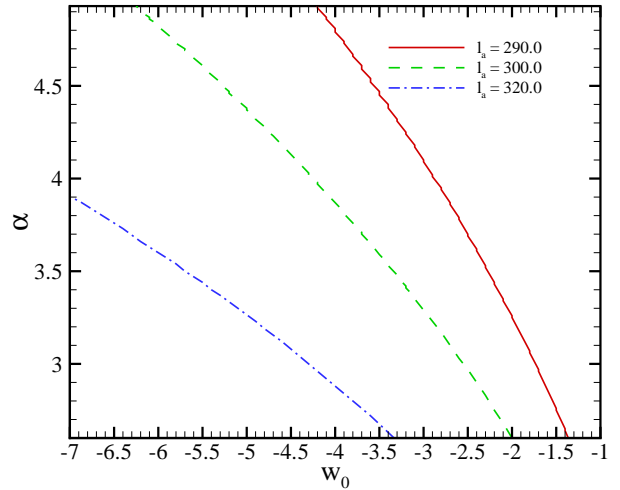} \narrowtext
\caption{Dependence of acoustic angular scale $l_A$ on $\alpha$
and $w_0$ for the three cases of $l_A=290$ (solid-line), $300$
(dashed-line) and $320$ (dashed-dotted line).} \label{lab}
\end{center}
\end{figure}
In order to compare the observed angular size of acoustic peak
with that of model, we use the shift parameter $R$ as
\cite{bond97}:
\begin{equation}
\label{shift} R=
\sqrt{\Omega_m}\int_0^{z_{dec}}\frac{dz}{E(z;\alpha,w_0)},
\end{equation}
where $E(z;\alpha,w_0)=H(z;\alpha,w_0)/H_0$. The shift parameter
is proportional to the size of acoustic peak to that of flat
pure-CDM, $\Lambda=0$ model, ($R\propto \theta_{A}/\theta_{A}^{\it
flat}$). The observational result of CMB experiments correspond a
shift parameter of $R=1.716\pm0.062$ (given by WMAP, CBI, ACBAR)
\cite{spe03,pearson03}. One of the advantages of using the
parameter $R$ is that it is independent of Hubble constant.

Recently detected size of baryonic peak in the SDSS is the third
observational data for our analysis.  The correlation function of
46,748 {\it Luminous Red Galaxies} (LRG) from the SDSS shows a
well detected baryonic peak around $100$ Mpc $h^{-1}$. This peak
has an excellent match to the predicted shape and the location of
the imprint of the recombination-epoch acoustic oscillation on the
low-redshift clustering matter \cite{eisenstein05}. For a flat
universe we can construct the parameter $A$ as follows:
\begin{equation} \label{lss1} A =
\sqrt{\Omega_m}E(z_1;\alpha,w_0)^{-1/3}
\times\left[\frac{1}{z_1}\int_0^{z_1}\frac{dz}{E(z;\alpha,w_0)}\right]^{2/3}.
\end{equation}
We use the robust constraint on the dark energy model using the
value of $A=0.469\pm0.017$ from the LRG observation at $z_1 =
0.35$ \cite{eisenstein05}.

In what follows we perform a combined analysis of SNIa, CMB and
SDSS to constrain the parameters of dark energy model by
minimizing the combined $\chi^2 = \chi^2_{\rm {SNIa}}+\chi^2_{{\rm
CMB}}+\chi^2_{{\rm SDSS}}$. The best values of the model
parameters from the fitting with the corresponding error bars from
the likelihood function marginalizing over the Hubble parameter in
the multidimensional parameter space results:
$\Omega_m=0.32_{-0.04}^{+0.03}$, $\alpha=1.60_{-0.90}^{+0.60}$ and
$w_0=-2.00_{-0.40}^{+0.80}$ at $1\sigma$ confidence level with
$\chi^2_{min}/N_{d.o.f}=1.13$. The Hubble parameter corresponds to
the minimum value of $\chi^2$ is $h=0.66$. Here we obtain an age
of $12.82$ Gyr for the universe. Table \ref{tab4} indicates the
best fit values for the cosmological parameters with one and two
$\sigma$ level of confidence.

\begin{table}
\begin{center}
\caption{\label{tab4} The best values for the parameters of
power-law dark energy model with the corresponding age for the
universe from the fitting with the SNIa, SNIa+CMB+SDSS and
SNIa+CMB+SDSS+LSS experiments at one and two $\sigma$ confidence
level.}
\begin{tabular}{|c|c|c|c|c|}
Observation & $\Omega_m$ & $\alpha$ & $w_0$ & age (Gyr) \\\hline
  &&&& \\
 & $0.45^{+0.09}_{-0.45}$&$1.00^{+1.00}_{-1.00}$ &$-2.60^{+1.80}_{-2.00}$ &  \\ 
 SNIa&&&&$13.19$\\
 & $0.45^{+0.13}_{-0.45}$&$1.00^{+2.00}_{-1.00}$ & $-2.60^{+1.90}_{-2.90}$ &
\\ &&&&\\ \hline
&&&& \\
SNIa+CMB& $0.32^{+0.03}_{-0.04}$&$1.60^{+0.60}_{-0.90}$ & $-2.00^{+0.80}_{-0.40}$& $12.82$ \\
+SDSS &&&& \\
&$0.32^{+0.05}_{-0.08}$&$1.60^{+1.40}_{-1.60}$  &
$-2.00^{+1.30}_{-1.30}$& \\
 &&&&\\ \hline
 &&&& \\
SNIa+CMB& $0.31^{+0.02}_{-0.04}$&$0.80^{+0.70}_{-0.30}$ & $-1.40^{+0.40}_{-0.65}$& $13.72$ \\
 SDSS+LSS&&&& \\
&$0.31^{+0.04}_{-0.06}$&$0.80^{+1.60}_{-0.80}$  & $-1.40^{+0.60}_{-1.10}$& \\
 &&&&\\
\end{tabular}
\end{center}
\end{table}

\section{Constraints by Large Scale structure }
\label{cstructure} So far we have only considered observations
related to the background evolution. In this section using the
linear approximation of structure formation we obtain the growth
index of structures and compare it with result of observations by
the $2$-degree Field Galaxy Redshift Survey ($2$dFGRS).

The continuity and Poisson equations for the density contrast
$\delta=\delta \rho/\bar{\rho}$ in the cosmic fluid provides the
evolution of density contrast in the linear approximation (i.e.
$\delta\ll 1 $)\cite{P93,B04} as:
\begin{equation}
\ddot{\delta}+2\frac{\dot{a}}{a}\dot{\delta} - \left( {v_s}^2
\nabla^2 +4\pi G \rho\right) \delta=0, \label{eq2}
\end{equation}
 where the dot denotes the derivative with respect to time.
The effect of dark energy in the evolution of the structures in
this equation enters through its influence on the expansion rate.
Here we assume that dark energy distributed uniformly as the
background fluid and it doesn't contribute in clustering of
matter. The validity of this linear Newtonian approach is
restricted to perturbations on the sub-horizon scales but large
enough where structure formation is still in the linear regime
\cite{P93,B04}. For the perturbations larger than the Jeans
length, $ \lambda_J= \pi^{1/2} v_s / \sqrt{G \rho} $, equation
(\ref{eq2}) for cold dark matter (CDM) reduces to:
\begin{equation} \ddot{\delta}+2 \frac{\dot{a}}{a} \dot{\delta}-4\pi G
\rho\delta=0 \label{eq3}
\end{equation}
The equation for the evolution of density contrast can be
re-written in terms of scale factor as:
\begin{equation}
\frac{d^2\delta}{da^2}+\frac{d\delta}{da}\left[\frac{\ddot{a}}{\dot{a}^2}+\frac{2H}{\dot{a}}\right]-
\frac{3H_0^2}{2\dot{a}^2a^3}\Omega_{m}\delta=0 \label{eq31}
\end{equation}
where the dot denotes the time derivative. Numerical solution of
equation (\ref{eq31}) in a FRW universe in the background of
power-law dark energy model is shown in Figure \ref{ev}. In the
CDM model, the density contrast $\delta$ grows linearly with the
scale factor, while we have a deviation from the linearity as soon
as dark energy begins to dominate. As larger $\alpha$ is, universe
enters the dark energy domination earlier (see Figure
\ref{fig_dr}) which results in a lesser growth of the density
contrast.
\begin{figure}[t]
\begin{center}
\epsfxsize=9.5truecm\epsfbox{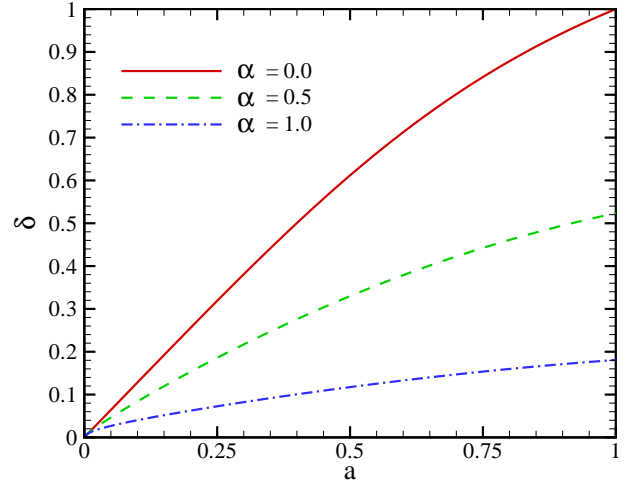} \narrowtext
\caption{Evolution of density contrast as a function of scale factor
for different values of $\alpha$ exponent in the flat universe with
$\Omega_m = 0.3$, $\Omega_{\lambda}=0.7$ and $w_0=-1.0$.} \label{ev}
\end{center}
\end{figure}

In the linear perturbation theory, the peculiar velocity field
$\bf{v}$ is determined by the density contrast \cite{P93,P80} as:
\begin{equation} {\bf v} ({\bf x})= H_0 \frac{f}{4\pi} \int \delta ({\bf y})
\frac{{\bf x}-{\bf y}}{\left| {\bf x}-{\bf y} \right|^3} d^3 {\bf
y}, \end{equation} where the growth index $f$ is defined by:
\begin{equation} f=\frac{d \ln \delta}{d \ln a},
 \label{eq5}
\end{equation}
and it is proportional to the ratio of the second term of equation
(\ref{eq3}) (friction) by the third (Poisson) term.

We use the evolution of the density contrast $\delta$ to compute
the growth index of structure $f$, which is an important quantity
for the interpretation of peculiar velocities of galaxies, as
discussed in \cite{P80,rah02} for the Newtonian and the
relativistic regime of structure formation. Replacing the density
contrast with the growth index in equation(\ref{eq31}) results the
evolution of growth index as:
\begin{eqnarray}
\label{index} &&\frac{df}{d\ln a}= \frac{3H_0^2}{2a^3}\Omega_m
-f^2 \\ \nonumber
&-&f\left[2-\frac{H_0^2}{2}[\frac{2}{H_0^2}+\frac{\Omega_m}{a^3}+
 \Omega_{\Lambda}(a)(1+3w(a))]\right]
\end{eqnarray}
Figure \ref{4} shows the numerical solution of (\ref{index}) in
terms of redshift.

\begin{figure}
\begin{center}
\epsfxsize=9.5truecm\epsfbox{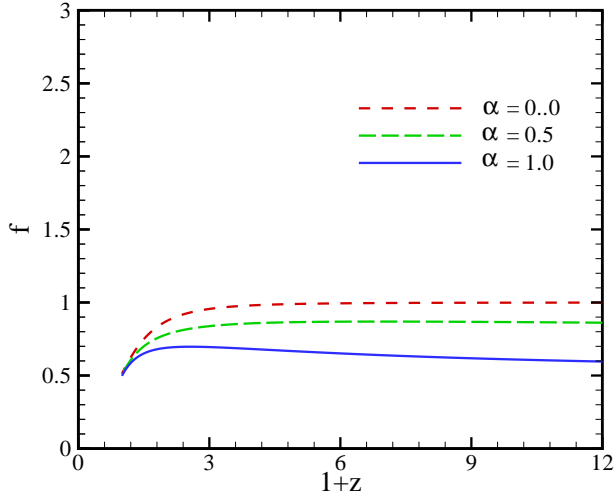} \narrowtext
\caption{Growth index versus redshift for different values of
$\alpha$. Here we take a typical values for the cosmological
parameters of $\Omega_m=0.3$ and $\Omega_{\lambda}=0.7$ and
$w_0=-1.0$.} \label{4}
\end{center}
\end{figure}

The observation of $220,000$ galaxies with the $2$dFGRS experiment
provides the numerical value of growth index \cite{eisenstein05}.
By measurements of two-point correlation function, the $2$dFGRS
team reported the redshift distortion parameter of $\beta = f/b
=0.49\pm0.09$ at $z=0.15$, where $b$ is the bias parameter
describes the difference in the distribution of galaxies and mass.
Verde et al. (2003) used the bispectrum of $2$dFGRS galaxies
\cite{ver01,la02} and obtained $b_{verde} =1.04\pm 0.11$ which
resulted $f= 0.51\pm0.10$. Now we fit the growth index at the
present time derived from the equation (\ref{index}) with the
observational value. This fitting gives a loss constraint to the
parameters of the model, so in order to have a better confinement
of the parameters, we combine this fitting with those of
SNIa$+$CMB$+$SDSS which has been discussed at the last section. We
perform the least square fitting by minimizing
$\chi^2=\chi_{\rm{SNIa}}^2+\chi_{\rm{CMB}}^2+\chi_{\rm{SDSS}}^2+\chi_{\rm{LSS}}^2$.
The best fit values with the corresponding error bars for the
model parameters are: $\Omega_m=0.31^{+0.02}_{-0.04}$,
$\alpha=0.80_{-0.30}^{+0.70}$ and $w_0=-1.40_{-0.65}^{+0.40}$ at
$1\sigma$ confidence level with $\chi^2_{min}/N_{d.o.f}=1.15$. The
error bars have been obtain through the likelihood functions
$(\mathcal{L}$$\propto e^{-\chi^2/2})$ marginalizing over the
nuisance parameter of $h$ \cite{press94}. The Hubble parameter
corresponds to the minimum value of $\chi^2$ is $h=0.65$. The
likelihood functions for the three cases of (i) fitting model with
Supernova data, (ii) combined analysis with the three experiments
of SNIa$+$CMB$+$SDSS and (iii) combining all four experiments of
SNIa$+$CMB$+$SDSS$+$LSS are shown in Figure \ref{mbhw}. The joint
confidence contours in the $(\Omega_m,w_0)$, $(\alpha,\Omega_m)$
and $(w_0,\alpha)$ planes also are shown in Figures \ref{jmw},
\ref{jbw} and \ref{jmb}.

Finally we do the consistency test, comparing the age of universe
derived from this model with the age of old stars and Old High
Redshift Galaxies (OHRG) in various redshifts.  Table \ref{tab4}
shows that the age of universe from the combined analysis of
SNIa$+$CMB$+$SDSS$+$LSS is $13.72$ Gyr which is in agreement with
the age of old stars \cite{carretta00}. Here we take three OHRG
for comparison with the power-law dark energy model, namely the
LBDS $53$W$091$, a $3.5$-Gyr old radio galaxy at $z=1.55$
\cite{dunlop96}, the LBDS $53$W$069$ a $4.0$-Gyr old radio galaxy
at $z=1.43$ \cite{dunlop99} and a quasar, APM $08279+5255$ at
$z=3.91$ with an age of $t=2.1_{-0.1}^{+0.9}$Gyr
\cite{hasinger02}. The later one has once again led to the "age
crisis". An interesting point about this quasar is that it cannot
be accommodated in the $\Lambda$CDM model \cite{jan06}. To
quantify the age-consistency test we introduce the expression
$\tau$ as:
\begin{equation}
 \tau=\frac{t(z;\alpha,w_0)}{t_{obs}} = \frac{t(z;\alpha,w_0)H_0}{t_{obs}H_0},
\end{equation}
where $t(z)$ is the age of universe, obtain from the equation
(\ref{age}) and $t_{obs}$ is an estimation for the age of old
cosmological object. In order to have a compatible age for the
universe we should have $\tau>1$. Table \ref{tab6} shows the value
of $\tau$ for three mentioned OHRG. We see that the parameters of
dark energy model from the SNIa and CMB observations don't provide
a compatible age for the universe, compare to the age of old
objects, while combination with the LSS data results a longer age
for the universe. Once again for the power-law dark energy model,
APM $08279+5255$ at $z=3.91$ has longer age than the universe.

\begin{table}
\begin{center}
\caption{\label{tab6} The value of $\tau$ for three high redshift
objects, using the parameters of the model derived from the
fitting with the observations.}
\begin{tabular}{|c|c|c|c|}
  Observation & LBDS $53$W$069$&LBDS $53$W$091$& APM  \\
&&& $08279+5255$ \\
  & $z=1.43$&$z=1.55$& $z=3.91$  \\ \hline
SNIa& $0.92$ & $0.97$& $0.59$ \\ \hline

SNIa+CMB & $0.83$&$0.88$&$0.50$ \\
 +SDSS& && \\
\hline
SNIa+CMB & $1.01$&$1.07$&$0.65$ \\
 +SDSS+LSS& && \\
\end{tabular}
\end{center}
\end{table}

\begin{figure}
\begin{center}
\epsfxsize=9.5truecm\epsfbox{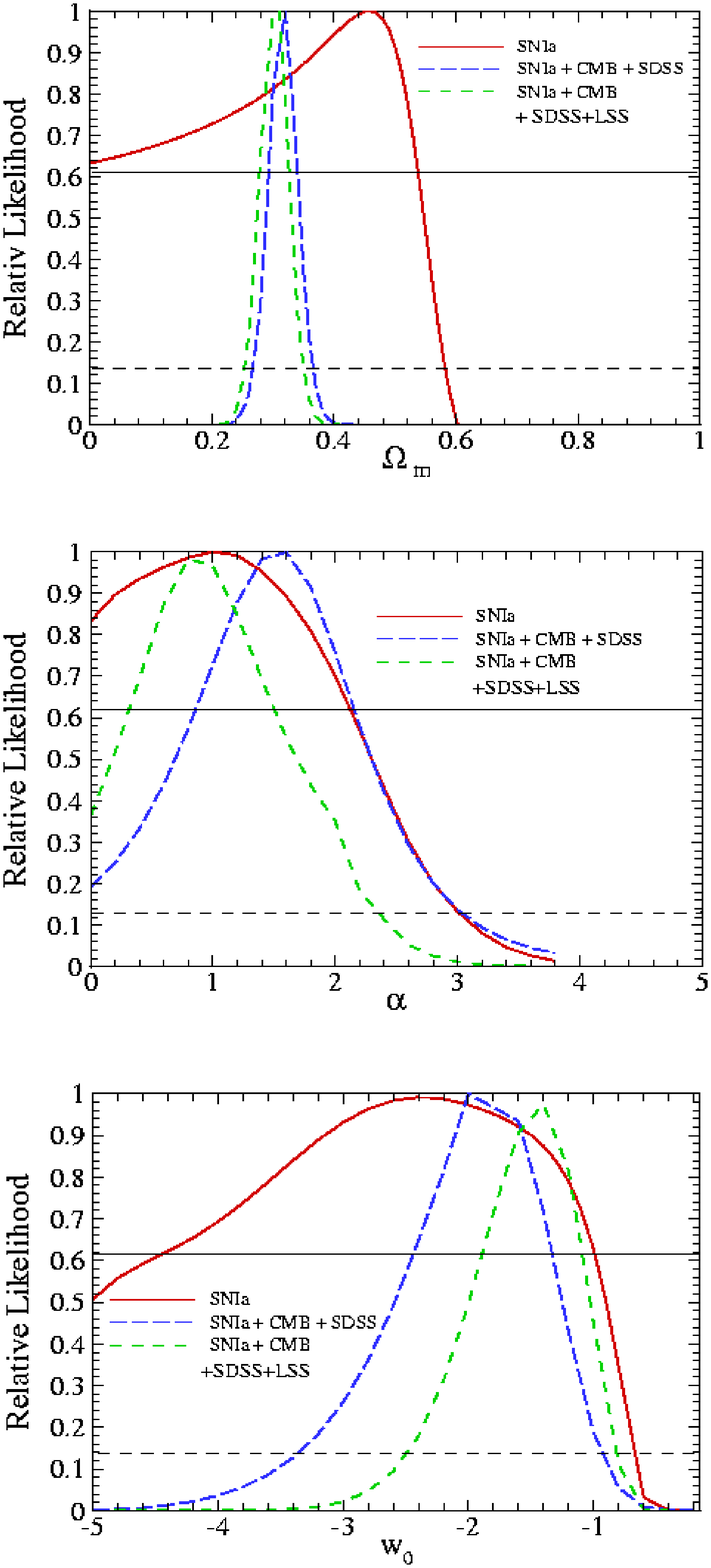} \narrowtext
\caption{Marginalized likelihood functions of three parameters of
dark energy model ($\Omega_M$, $\alpha$ and $w_0$). The solid line
corresponds to the likelihood function of fitting the model with
SNIa data, the long dashed-line with the joint SNIa$+$CMB$+$SDSS
data and dashed-line corresponds to SNIa$+$CMB$+$SDSS$+$LSS. The
intersections of the curves with the horizontal solid and dashed
lines give the bounds with $1\sigma$ and $2\sigma$ level of
confidence respectively.} \label{mbhw}
\end{center}
 \end{figure}


\begin{figure}[t]
\epsfxsize=9.0truecm\epsfbox{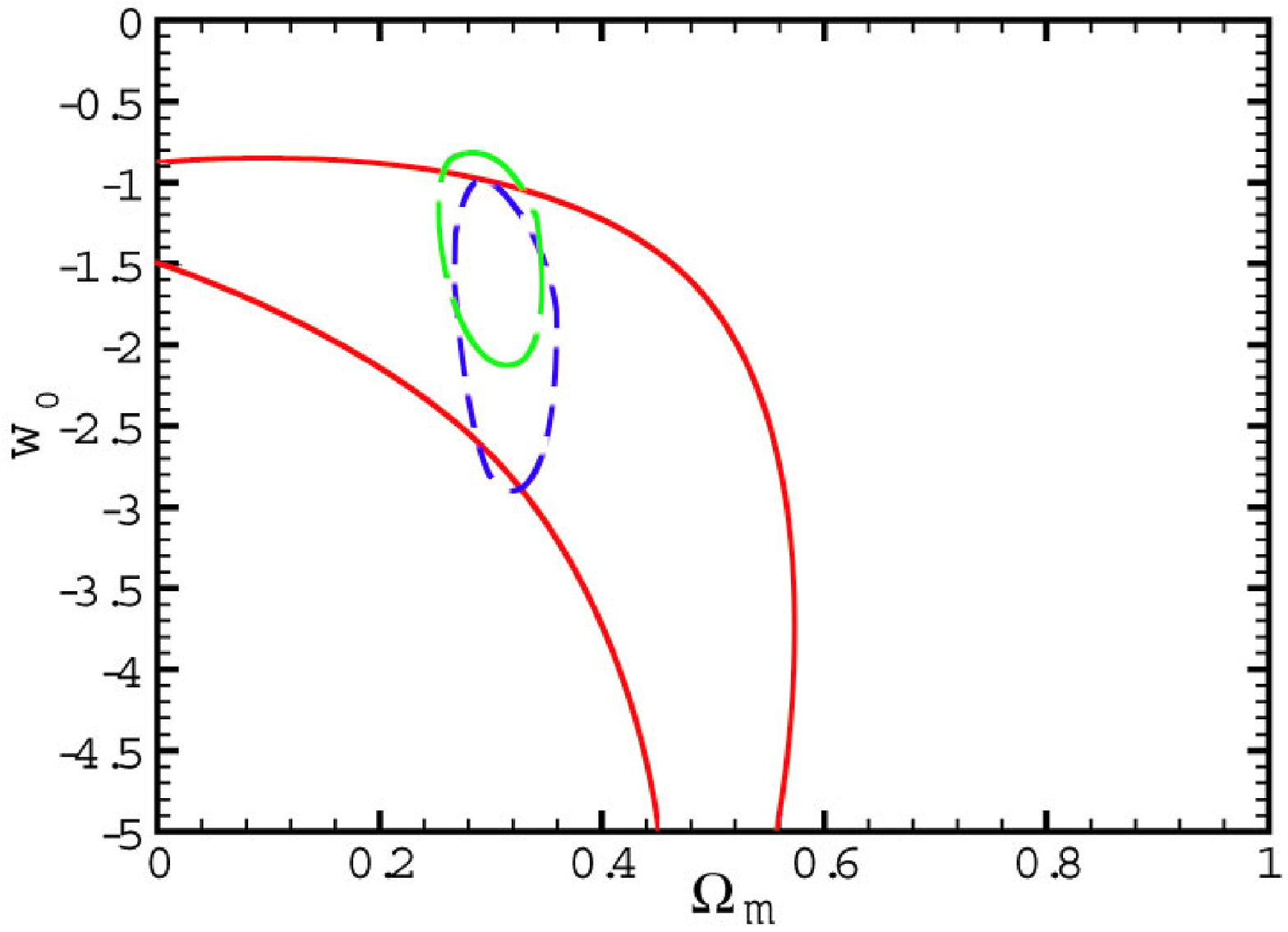} \narrowtext \caption{Joint
confidence intervals of $\Omega_m$ and $w_0$, fitting with SNIa
(solid line), SNIa$+$CMB$+$SDSS (dashed-line) and
SNIa$+$CMB$+$SDSS$+$LSS (long dashed-line) with $1\sigma$ level of
confidence.} \label{jmw}
 \end{figure}

\begin{figure}[t]
\epsfxsize=9.0truecm\epsfbox{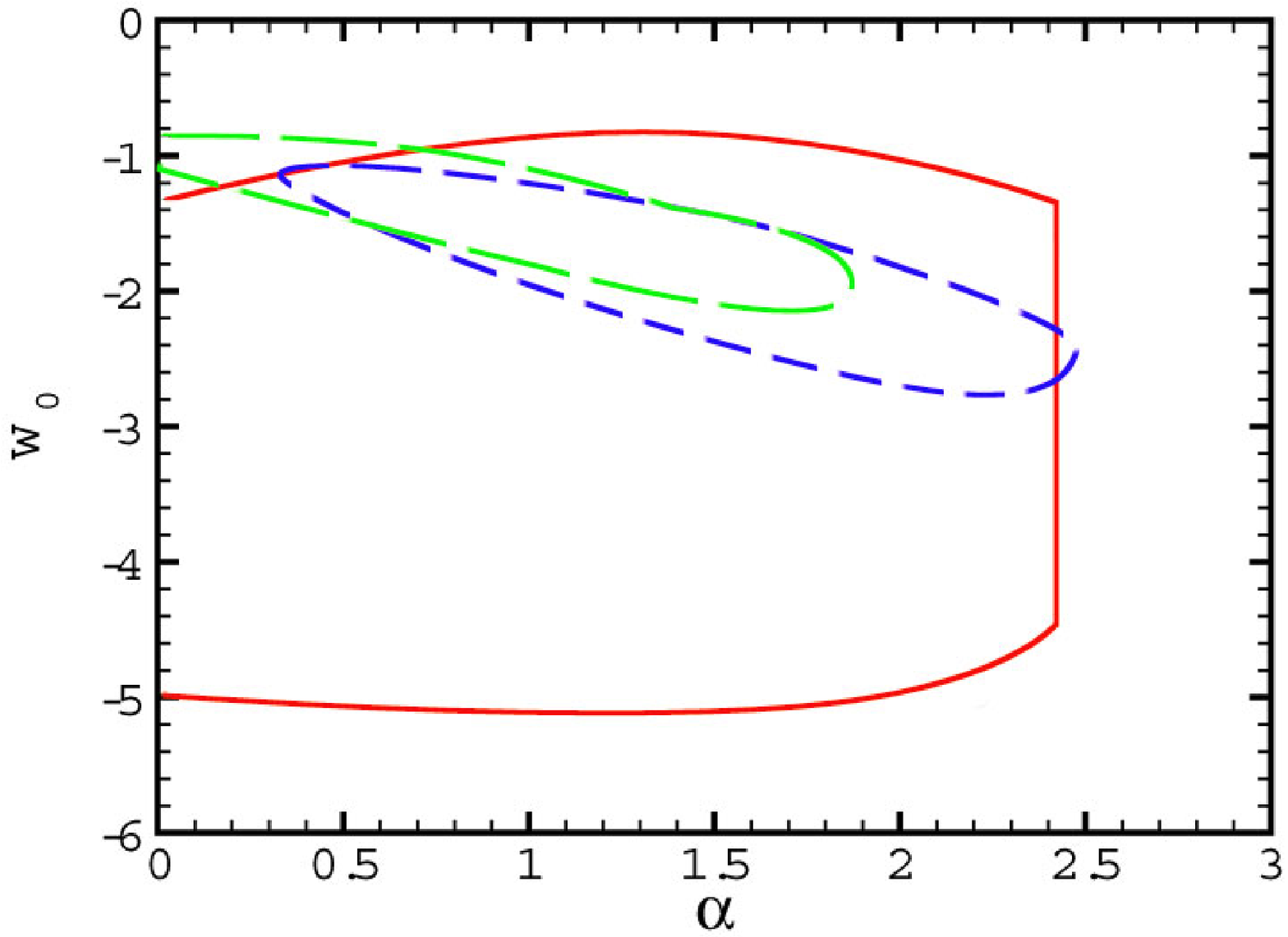} \narrowtext \caption{Joint
confidence intervals of $\alpha$ and $w_0$, fitting with the SNIa
(solid line), SNIa$+$CMB$+$SDSS (dashed-line) and
SNIa$+$CMB$+$SDSS$+$LSS (long dashed-line) with $1\sigma$ level of
confidence.} \label{jbw}
 \end{figure}
\begin{figure}
\epsfxsize=9.0truecm\epsfbox{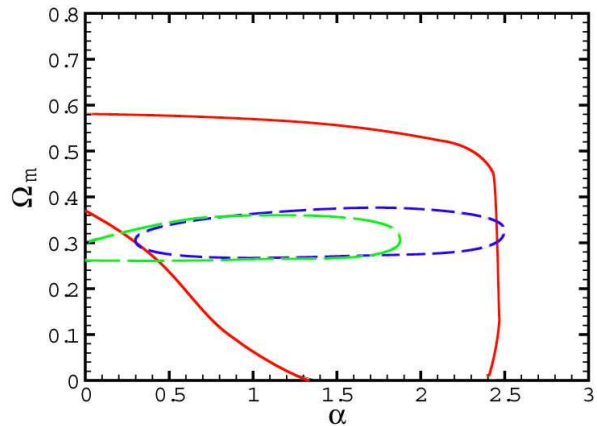} \narrowtext \caption{Joint
confidence intervals of $\Omega_m$ and $\alpha$, fitting with the
SNIa (solid line), SNIa$+$CMB$+$SDSS (dashed-line) and
SNIa$+$CMB$+$SDSS$+$LSS (long dashed-line) with $1\sigma$ level of
confidence.} \label{jmb}
 \end{figure}

\section{conclusion}
\label{conc} We proposed a power-law parameterized quintessence
model with the mean-equation of state of $\bar{w}(z) =
w_0a^{\alpha}$. An exponential potential of scalar field is
proposed for generating this type of dark energy model. The effect
of this model on the age of universe, radial comoving distance,
comoving volume element and the variation of apparent size of
objects with the redshift (Alcock-Paczynski test) have been
studied. In order to constrain the parameters of model we fit our
model with the Gold sample SNIa data, CMB shift parameter,
location of baryonic acoustic peak observed by SDSS and large
scale structure data by $2$dFGRS. The best parameters from the
fitting obtained as: $h=0.65$, $\Omega_m=0.31^{+0.02}_{-0.04}$,
$\alpha=0.80_{-0.30}^{+0.70}$ and $w_0=-1.40_{-0.65}^{+0.40}$ at
$1\sigma$ confidence level with $\chi^2_{min}/N_{d.o.f}=1.15$. The
best fit for the equation of state at the present time provides
that $w_0<-1$, which violates the strong energy condition in
general relativity. Furthermore for $w_0<-1$ the kinetic term of
scalar field in the Lagrangian is negative \cite{car03,ha73}
(theoretical attempts for $w <-1$ can be found at
~\cite{Caldwell:1999ew,Schulz:2001yx,parker,frampton,Ahmed:2002mj,CHT}).

We also did the age test, comparing the age of old stars and old
high redshift galaxies with the age derived from the power-law
dark energy model. From the best fit parameters of the model we
obtained an age of $13.72$ Gyr for the universe which is in
agreement with the age of old stars. We also chose two high
redshift radio galaxies at $z=1.55$ and $z=1.43$ with a quasar at
$z=3.91$. The two first objects were consistent with the age of
universe by means that there were younger than the age of universe
while the later one was older than the age of universe.


\end{document}